\documentclass[12pt]{article}
\usepackage{amsmath}
\usepackage{amsfonts}
\usepackage{graphicx}
\usepackage{amssymb}
\usepackage[unicode,bookmarks,bookmarksopen,bookmarksopenlevel=2,colorlinks,linkcolor=blue,citecolor=green]{hyperref}

\input{tcilatex}
\begin{document}

\title{\textbf{Unfolding of singularities and differential equations}}
\author{B. Konopelchenko \\
Dipartimento di Fisica, Universita del Salento and INFN,\\
sezione di Lecce, 73100 Lecce, Italy}
\maketitle

\begin{abstract}
Interrelation between Thom's catastrophes and differential equations
revisited. It is shown that versal deformations of critical points for
singularities of A,D,E type are described by the systems of Hamilton-Jacobi
type equations. For particular nonversal unfoldings the corresponding
equations are equivalent to the integrable two-component hydrodynamic type
systems \ like classical shallow water equation, dispersionless Toda system
and others. Peculiarity of such integrable systems is that the generating
functions for the corresponding hierarchies, which obey
Euler-Poisson-Darboux equation, contain information about normal forms of
higher order and higher corank singularities.
\end{abstract}

\tableofcontents

\ \ \ \ \ \ \ \ \ \ \ \ \ \ \ \ \ \ \ \ \ \ \ 

\section{Introduction}

\bigskip

Connection between catastrophe theory ( see e.g. \cite{Thom,
Lu,poston,AGV1,AGV2}) and theory of singularities for differential equations
has been studied for more than fourty years. It was R.Thom himself \cite%
{Thom} to observe that the singularities developed by solutions of the wave
equation are related to the unfolding of singularities of functions. In 1968
E. Zeeman noted \cite{Zeeman} that the breaking waves phenomenon (gradient
catastrophe) is associated with the hyperbolic umbilic catastrophe. Within
the formalism of Lagrangian submanifolds the above interrelation has been
studied by Guckenheimer \cite{Guck}, Arnold \cite{Arnold2,Arnold3} and other
authors ( see e.g. \cite%
{Janich,Berry,Givental1,Malikov,Zakal,Lychagin,Rakhimov} and references
therein). Some aspects of the tangent fields for singularity unfoldings in
connection with deformations of complex spaces have been considered in \cite%
{Palamod} . In topological quantum field theory \cite{Witten1} unfoldings of
singularities and corresponding Witten-Dijkgraaf-Verlinde-Verlinde (WDVV)
equations have appeared within the study of the deformed Ginzburg-Landau
models \cite{Witten2,DVV}. \ A connection of the topological field theory
and Saito's approach in singularity theory \cite{Saito} with Frobenius
manifolds \cite{Dubr1}and associated WDVV and Hirora equations has been
discussed in \cite%
{Dubr1,Dubr2,Krichever,Hert,Givental2,Giv-milanov,Fren-Giv-Milan}. It is
well established nowadays that the base spaces of versal unfoldings of
singularities may carry rich family of algebraic and geometrical structures
including various types of differential equations.

Within different studies an appearance of normal forms of singularities has
been observed recently in the papers \cite{Kodama-Kon,
Dubr3,Dubr4,Dubr-Grava-Klein,Kon-Alonso-Med1,Kon-Alonso-Med2}. \ In \cite%
{Dubr3,Dubr4,Dubr-Grava-Klein} cusp and umbilic catastrophes have arozen
within the study of the behaviour of the full dispersive integrable
equations \ near the points of gradient catastrophe. \ In contrast, an
analysis performed in \cite{Kodama-Kon,Kon-Alonso-Med1,Kon-Alonso-Med2}
deals directly with the dispersionless integrable systems of certain class.
In the papers \cite{Kon-Alonso-Med1,Kon-Alonso-Med2,Kon-Alonso-Med3} it was
shown that the hodograph equations for these systems are nothing else than
the equations defining critical points of the functions $W$ which obey
Euler-Poisson-Darboux equations. Such functions $W$ \ have several important
properties: they are the generating functions for the whole integrable
hierarchies, they drastically simplify an analysis of the singular sectors
for these hierarchies. But, perhaps, more interesting fact is that they
contain information about the normal forms of singularities of functions
from the catastrophe theory. The observations made in the papers \cite%
{Kodama-Kon,Kon-Alonso-Med1,Kon-Alonso-Med2} deal with the singularities of $%
A_{n}$ type \ and umbilic type.

In the present paper we will discuss one particular aspect of the relation
between the singularity theory and differential equations. Let the function $%
F(x;t)=F_{0}(x)+\sum_{k=0}^{n}e_{k}(x)t_{k}$ with certain functions $%
e_{k}(x) $ and deformation parameters $t_{k}$ defines the unfolding of the
singularity with the normal form given by the function $F_{0}(x)$ of m
variables $x_{1},...,x_{m}$. We will be interested in differential equations
which govern the dependence of the critical points $u_{i},i=1,...,m$ for the
function $F(x;t)$ on deformation parameters. We will show, using an
elementary technique and \ addressing mainly to nonexperts, that

1. For the versal deformations of the A,D,E singularities ( $m=1$ or 2) the
dependence of their critical points on $t_{k}$ outside the catastrophe sets
is described by the systems of mn differential equations of the first order.
These systems imply that $u_{i}=\varphi _{t_{i}}$ and the function \ $%
\varphi $ obeys the systems of Hamilton-Jacobi type equations. For $A_{n}$
singularities these equations are of the form

\begin{equation*}
\varphi _{t_{k}}=(\varphi _{t_{1}})^{k},\quad k=2,...,n-1
\end{equation*}%
or

\begin{equation*}
u_{t_{k}}=(u^{k})_{t_{1}},\quad k=2,...,n-1,
\end{equation*}%
which are the\ Burgers-Hopf and n-3 higher Burgers-Hopf equations. For $%
D_{n},E_{6},E_{7},E_{8}$ cases (m=2) the corresponding equations are

\begin{equation*}
\varphi _{t_{l}}=e_{l}(\varphi _{t_{1}},\varphi _{t_{2}}),\quad l=3,...,\mu
-1
\end{equation*}%
or

\begin{equation*}
u_{t_{l}}=\frac{\partial e_{l}(u,v)}{\partial t_{1}},\quad v_{t_{l}}=\frac{%
\partial e_{l}(u,v)}{\partial t_{2}},\quad l=3,...,\mu -1
\end{equation*}%
where $u=u_{1},v=u_{2}$ and $e_{l}(x,y)$ are elements of a basis of the
local algebra $Q_{F_{0}}$ for critical points of the dimension $\mu $. In
all these cases for the versal deformations

\begin{equation}
F(u;dt)=d\varphi .
\end{equation}

2. For the particular nonversal unfolding of the umbilic singularities, for
which $F(x,y;t)=F_{0}(x,y)$ $+$ $t_{3}F_{0x}$ $+t_{1}x$ and the constraint $%
(F_{xx}-\delta F_{yy})(u,v)=0$ with $\delta =\pm 1$ is verified, the
critical points $u(t)$ and $v(t)$ obey integrable systems of hydrodynamic
type of the form

\begin{equation}
\left( 
\begin{array}{c}
u \\ 
v%
\end{array}%
\right) _{t_{3}}=2\left( 
\begin{array}{cc}
\alpha u, & v \\ 
\delta v, & \alpha u%
\end{array}%
\right) \left( 
\begin{array}{c}
u \\ 
v%
\end{array}%
\right) _{t_{1}}.
\end{equation}%
\ \ For certain values of $\alpha $ this system represents the well-known
integrable systems like \ the classical shallow waver equation ($\alpha =2$)
and dispersionless Toda system ($\alpha =0$). In this case the catastrophe
set defined by the equation $\Delta \div (F_{xx}F_{yy}-F_{xy}^{2})(u,v)=0$
decomposes in two components. For such integrable systems the corresponding
hodograph equations coincide with the equations for critical points\ $u$ and 
$v$ of the functions $W$ obeying the Euler-Poisson-Darboux equations. For
this unfolding and other nonversal unfoldings of such type the relation (1)
is valid too.

3. It is observed that the generating functions $W(x;t)=\sum_{k\geq
1}t_{k}W_{k}(x)$ for the integrable n-component hydrodynamic type systems
contain information about normal forms of singularities of higher orders and
coranks. \ It is shown that in the two-component case, namely, for the
system (2)

\begin{equation*}
W_{3}\thicksim x^{3}\pm 3xy^{2},\quad W_{4}\thicksim x^{4}+dx^{2}y^{2}+y^{4},
\end{equation*}%
i.e. the functions $W_{3}$ and $W_{4}$ are proportional to the normal forms
of $D_{4}$ and $X_{9}$ singularities of corank two, respectively. In the
three-component case

\begin{equation*}
W_{3}\thicksim x^{3}+y^{3}+z^{3}+dxyz,
\end{equation*}%
that is the normal form of the unimodular singularity $P_{8}$ (or $T_{3,3,3}$%
) of corank three. Dependence of the critical points for this singularity on
deformation parameters is governed by the dispersionless three-component
coupled KdV equation.

\bigskip

\section{Thom's catastrophes}

\bigskip

First, we recall briefly the basic facts about unfolding of singularities
(see \cite{Thom, Lu,poston,AGV1,AGV2}). Let the normal form of the
singularity of the corank m is given by the function $F_{0}(x_{1},...,x_{m})$
. An unfolding ( deformation) $F(x;\lambda )$ of the function $F_{0}$ is a
germ of a smooth function $F:R^{m}\times R^{n}\rightarrow R$ at the point
(0,0) such that $F(x;0)=F_{0}(x)$. The space $R^{n}$ of the second argument
of F is called the base space of unfolding and $\lambda _{1},...,\lambda
_{n} $ are called the parameters of unfolding. Dimension of the base space
for different deformations of the same function can be different. Critical
set of $F(x;\lambda )$ is given by the solutions of the equations \ $%
F_{x_{i}}(u;\lambda )\div \frac{\partial F}{\partial x_{i}}(u;\lambda
)=0,i=1,...,m$ and the catastophe set is defined by vanishing of the Hessian 
$\left\vert F_{x_{i}x_{k}}\right\vert (u)=0.$ The tangent space of the orbit
of the germ $F_{0}(x)$ is its gradient ( Jacobian) ideal $I_{J}=($ $%
F_{0x_{1}},...,F_{0x_{m}})$ which consists of all germs of the functions of
the form $\sum_{i=1}^{i=m}h_{i}(x)F_{0x_{i}}$. Local algebra $Q_{F_{0}}$ of
the critical point is the quotient $R[x]/I_{J}$. It's dimension $\mu $ is
called the Milnor number ( or multiplicity) of the critical point. If the
functions $e_{0}=1,e_{1}(x),...,e_{\mu -1}(x)$ form a basis in $Q_{F_{0}}$,
then a versal\ ( R-versal) unfolding of $F_{0}(x)$ can be represented in the
form

\quad 
\begin{equation}
F(x;t)=F_{0}(x)+\sum_{k=0}^{\mu -1}e_{k}(x)t_{k}.
\end{equation}%
Here $t_{k}$ are the deformation parameters and hence the initial velocities
of deformation $V_{k}\div F_{t_{k}}(x;t)|_{t=0}=e_{k}(x),k=0,1,...,\mu -1.$
Nonversal unfoldings of \ $F_{0}(x)$ may depend on less than $\mu $
parameters or have other forms.

\ Let us begin with the celebrated seven Thom's catastrophes. The
corresponding functions $F_{0}$ and their versal unfoldings are (see \cite%
{Thom, Lu,poston,AGV1,AGV2}): for corank one $A_{n}$ type catastrophes (
fold, cusp, swallow tail, butterfly)

\bigskip

\begin{equation}
F_{n}^{A}(x;t)=x^{n+1}+\sum_{k=0}^{n-1}t_{k}x^{k},\quad n=2,3,4,5,
\end{equation}%
for corank two hyperbolic and elliptic umbilic catastrophes ( $D_{4}^{\pm }$)

\begin{equation}
F_{4}^{\pm }(x,y;t)=x^{3}\pm 3xy^{2}+t_{3}(x^{2}\mp
y^{2})+t_{2}y+t_{1}x+t_{0},
\end{equation}%
and for the parabolic umbilic catastrophe $D_{5}$

\begin{equation}
F_{5}(x,y;t)=x^{2}y+y^{4}+t_{4}y^{2}+t_{3}x^{2}+t_{2}y+t_{1}x+t_{0}.
\end{equation}

We begin with the $A_{n}$ type singularities. Critical points $u$ of the
functions (4) are defined by the equation

\begin{equation}
F_{nu}^{A}\div F_{nx}^{A}(u;t)=(n+1)u^{n}+\sum_{k=1}^{n-1}kt_{k}u^{k-1}=0
\end{equation}%
where $F_{x}$ etc denotes the derivative w.r.t. $x$. \ Calculating the
differential of (7), one gets

\begin{equation}
F_{n\unit{u}u}^{A}du+\sum_{k=1}^{n-1}ku^{k-1}dt_{k}=\sum_{k=1}^{n-1}(F_{n%
\unit{u}u}^{A}u_{t_{k}}+ku^{k-1})dt_{k}=0.
\end{equation}%
Hence, \ the critical points $u(t)$ obey the system of equations

\begin{equation}
F_{n\unit{u}u}^{A}u_{t_{k}}+ku^{k-1}=0,\quad k=1,2,...,n-1.\ 
\end{equation}%
Outside the catastrophe sets where $F_{n\unit{u}u}^{A}\neq 0$ one has the
system

\begin{equation}
u_{t_{k}}=-\frac{ku^{k-1}}{F_{n\unit{u}u}^{A}},\quad k=1,2,...,n-1.
\end{equation}%
On the catastrophe sets defined by the condition $F_{n\unit{u}u}^{A}=0$ the
velocities $u_{t_{k}}$ of variations of the positions of the critical point
becomes unbounded which is a standard manifestation of the catasrophe.

For \ $n=2$ one has a single equation while at $n\geq 3$ the system (10) is
equivalent to the system

\begin{equation}
u_{t_{1}}=-\frac{1}{F_{n\unit{u}u}^{A}},\quad \
u_{t_{k}}=(u^{k})_{t_{1}},\quad k=2,3,...,n-1
\end{equation}%
Equations (11) imply that there exists a function $\varphi $ such that $%
u=\varphi _{t_{1}}$ and that the n-2 last equations (11) take the form

\begin{equation}
\varphi _{t_{k}}=(\varphi _{t_{1}})^{k},\quad k=2,3,...,n-1.
\end{equation}%
As a consequence, evaluating the infinitesimal variation of the function (4)
on the critical set, one obtains

\begin{equation}
\delta F_{n}^{A}(u)\div
F_{n}^{A}(u;t+dt)-F_{n}^{A}(u;t)=F_{n}^{A}(u;dt)=d\varphi .
\end{equation}

Equation (11) \ with $k=2$ is the well-known Burgers-Hopf (BH) equation (see
e.g.\cite{Whitham} ) while equations (11) at $k\geq 3$ represent $n-3$
higher flows commuting with them. In the form (12) this system is the family
of Hamilton-Jacobi type equations. \ The fact that the deformations of the
critical points for the first four Thom's catastrophes are governed by the
BH and higher BH equations has been discussed in \cite{Kodama-Kon} and \cite%
{Givental2}.

Umbilic type singularities can be considered simultaneously. Indeed, let us
take the function

\begin{equation}
F(x,y;t)=F_{0}(x,y;t)+t_{4}y^{2}+t_{3}x^{2}+t_{2}y+t_{1}x+t_{0}.
\end{equation}%
For $D_{4}^{+}$ singularity $F_{0}=x^{3}+3xy^{2}$ and $t_{4}+t_{3}=0$.
\bigskip For $D_{4}^{-}$ singularity $F_{0}=x^{3}-3xy^{2}$ and $%
t_{4}-t_{3}=0 $ and in $D_{5}$ case $F_{0}=x^{2}y+y^{4}$. Critical set is
defined by the equations

\begin{equation}
F_{x}(u,v;t)=F_{0x}(u,v)+2t_{3}x+t_{1}=0,
\end{equation}

\begin{equation}
F_{y}(u,v;t)=F_{0y}(u,v)+2t_{4}y+t_{2}=0.
\end{equation}%
Calculating differentials of these equations, taking into account that in
the critical set $du=\sum_{k=1}^{k=4}u_{t_{k}}dt_{k},$ $dv=%
\sum_{k=1}^{k=4}v_{t_{k}}dt_{k}$ and solving systems of linear equations,
one gets

\begin{equation}
u_{t_{1}}=-\frac{F_{vv}}{\Delta },\quad v_{t_{1}}=\frac{F_{uv}}{\Delta },
\end{equation}

\begin{equation}
u_{t_{2}}=\frac{F_{uv}}{\Delta },\quad v_{t_{2}}=-\frac{F_{\unit{u}u}}{%
\Delta },
\end{equation}

\begin{equation}
u_{t_{3}}=-2u\frac{F_{vv}}{\Delta },\quad v_{t_{3}}=2u\frac{F_{uv}}{\Delta },
\end{equation}

\begin{equation}
u_{t_{4}}=2v\frac{F_{uv}}{\Delta },\quad v_{t_{4}}=-2v\frac{F_{\unit{u}u}}{%
\Delta }
\end{equation}%
where \ $F_{\unit{u}u}=F_{xx}(u,v)$ etc and $\Delta =F_{\unit{u}%
u}F_{vv}-(F_{uv})^{2}$. \ These equations govern the dependence of the
critical points on deformation parameters. On the catastrophe set $\Delta =0$
derivatives of $u$ and $v$ become infinite and one observes a very fast (
catastrophic) change of positions and number of critical points.

The system of equations (17)-(20) can be represented in different equivalent
forms. For example, it implies that

\begin{equation}
u_{t_{2}}=v_{t_{1}},
\end{equation}%
and

\begin{equation}
u_{t_{3}}=(u^{2})_{t_{1},}\quad v_{t_{3}}=(u^{2})_{t_{2}},
\end{equation}

\begin{equation}
u_{t_{4}}=(v^{2})_{t_{1},}\quad v_{t_{4}}=(v^{2})_{t_{2}.}
\end{equation}%
Hence, there exists a function $\varphi $ such that $u=\varphi
_{t_{1}},v=\varphi _{t_{2}}$ and equations (22) and (23) are reduced to two
equations

\begin{equation}
\varphi _{t_{3}}=(\varphi _{t_{1}})^{2},\quad \varphi _{t_{4}}=(\varphi
_{t_{2}})^{2}.
\end{equation}%
It is easy to see that the system (17)-(20) is equivalent to the following
system for the function $\varphi $

\begin{equation}
\varphi _{t_{1}t_{1}}=-\frac{F_{vv}}{\Delta },\quad \varphi _{t_{1}t_{2}}=%
\frac{F_{uv}}{\Delta },\quad \varphi _{t_{2}t_{2}}=-\frac{F_{\unit{u}u}}{%
\Delta },
\end{equation}

\begin{equation}
u=\varphi _{t_{1}},\quad v=\varphi _{t_{2}},
\end{equation}

\begin{equation}
\varphi _{t_{3}}=(\varphi _{t_{1}})^{2},\quad \varphi _{t_{4}}=(\varphi
_{t_{2}})^{2}.
\end{equation}%
As the consequence of equations (26), (27) one has

\begin{equation}
\delta F(u,v)\div F(u,v;t+dt)-F(u,v;t)=F(u,v;dt)=\sum_{k=0}^{k=4}\varphi
_{t_{k}}dt_{k}=d\varphi
\end{equation}%
with $\varphi _{t_{0}}=1$.

Equations (17)-(27) describe dependence of critical points of $D_{5}$
singularity on deformation parameters. \ To get the corresponding equations
for $D_{4}^{\pm }$ singularities one can pass in equations (17)-(27) to new
variables $t_{\pm }=\frac{1}{2}(t_{4}\pm t_{3})$ and then impose the
constraints $t_{+}=0$ or $t_{-}=0$ or use directly the relation (28).
Indeed, the functions $F^{\pm }$ for $D_{4}^{\pm }$ singularities are $%
F^{\pm }=F(x,y;t)|_{t_{4}=\mp t_{3}}$. So,

\begin{equation}
d\varphi ^{\pm }=\sum_{k=0}^{k=3}\varphi _{t_{k}}^{\pm }dt_{k}=d\varphi
|_{_{t_{4}=\mp t_{3}}}=((\varphi _{t_{1}}^{\pm })^{2}\mp (\varphi
_{t_{2}}^{\pm })^{2})dt_{3}+\varphi _{t_{2}}^{\pm }dt_{2}+\varphi
_{t_{1}}^{\pm }dt_{1}+dt_{0}.
\end{equation}%
Hence, for the $D^{\pm }$ \ singularities one has the equation%
\begin{equation}
\varphi _{t_{3}}^{\pm }=(\varphi _{t_{1}}^{\pm })^{2}\mp (\varphi
_{t_{2}}^{\pm })^{2}.
\end{equation}%
plus the corresponding equations (25), (26). \ In terms of $u=\varphi
_{t_{1}}^{\pm }$ and $v=$ $\varphi _{t_{2}}^{\pm }$ equation (30) assumes
the form of the 2+1-dimensional hydrodynamical type system

\begin{eqnarray}
u_{t_{3}} &=&(u^{2}\mp v^{2})_{t_{1}},  \notag \\
v_{t_{3}} &=&(u^{2}\mp v^{2})_{t_{2}}.
\end{eqnarray}

\ The catastrophe sets $\Delta =0$ for the umbilic singularities discussed
above are those subsets of variables $t_{k}$ for which solutions of
considered systems of differential equations exhibit gradient catastrophe $%
u_{t_{k}},v_{t_{k}}\rightarrow \infty $.

\bigskip\ 

\section{A,D,E singularities}

In his seminal paper \cite{Arnold2} Arnold proved that the list of functions
with simple (without moduli) degenerate critical points consists of two
infinite series $A_{n}(n\geq 1),D_{n}(n\geq 4)$ and three exceptional cases $%
E_{6},E_{7},E_{8}$. The corresponding normal forms and versal unfoldings are
as follows:

\begin{equation}
A_{n}:\quad F(x;t)=x^{n+1}+t_{n-1}x^{n-1}+t_{n-2}x^{n-2}+...+t_{1}x+t_{0},
\end{equation}

\begin{equation}
D_{n}:\quad F(x,y;t)=x^{2}y\pm y^{n-1}+t_{n-1}y^{n-2}+...t_{2}y+t_{1}x+t_{0,}
\end{equation}

\begin{equation}
E_{6}:\quad F(x,y;t)=x^{3}\pm
y^{4}+t_{5}xy^{2}+t_{4}xy+t_{3}y^{2}+t_{2}y+t_{1}x+t_{0},
\end{equation}

\begin{equation}
E_{7}:\quad
F(x,y;t)=x^{3}+xy^{3}+t_{6}xy+t_{5}y^{4}+t_{4}y^{3}+t_{3}y^{2}+t_{2}y+t_{1}x+t_{0},
\end{equation}

\begin{equation}
E_{8}:\quad
F(x,y;t)=x^{3}+y^{5}+t_{7}xy^{3}+t_{6}xy^{2}+t_{5}xy+t_{4}y^{3}+t_{3}y^{2}+t_{2}y+t_{1}x+t_{0}.
\end{equation}

Equations governing the dependence of the critical points of $A_{n}$ type
has been found, in fact, in the previous section. It is sufficient to extend
the formulas \ (7)-(13) to arbitrary n. Thus, in this case (for $n\geq 3$)
one has \ the system of equations

\begin{equation}
u_{t_{k}}=(u^{k})_{t_{1}},\quad k=2,3,...,n-1
\end{equation}%
or

\begin{equation}
\varphi _{t_{k}}=(\varphi _{t_{1}})^{k},\quad k=2,3,...,n-1
\end{equation}%
and $F(u;dt)=d\varphi $. The system (37) contains the BH equation together
with its $n-3$ higher flows. Thus, the system of differential equations
governing versal deformations of the critical points for the entire $A_{n}$
series with unbounded n represents the whole infinite BH hierarchy \ (see 
\cite{Kodama-Kon} and \cite{Givental2}).

In order to consider D and E cases let us write the functions (33)-(36) as

\begin{equation}
F(x,y;t)=F_{0}(x,y)+\sum_{k=0}^{\mu -1}e_{k}(x,y)t_{k}
\end{equation}%
with corresponding $F_{0},e_{k}$ and $\mu $. Critical points are defined by
the system

\begin{equation}
F_{u}=F_{0u}+\sum_{k=0}^{\mu -1}e_{ku}t_{k}=0,
\end{equation}

\begin{equation}
F_{v}=F_{0v}+\sum_{k=0}^{\mu -1}e_{kv}t_{k}=0.
\end{equation}%
Differentiating these equations w.r.t. $t_{l}$, one gets the system

\begin{equation}
\left( 
\begin{array}{cc}
F_{\unit{u}u} & F_{uv} \\ 
F_{uv} & F_{vv}%
\end{array}%
\right) \left( 
\begin{array}{c}
u_{t_{l}} \\ 
v_{t_{l}}%
\end{array}%
\right) =-\left( 
\begin{array}{c}
e_{lu} \\ 
e_{lv}%
\end{array}%
\right) ,\quad l=1,...,\mu -1
\end{equation}%
and consequently

\begin{equation}
u_{t_{l}}=-\frac{1}{\Delta }\left\vert 
\begin{array}{cc}
e_{lu} & F_{uv} \\ 
e_{lv} & F_{vv}%
\end{array}%
\right\vert ,\quad v_{t_{l}}=-\frac{1}{\Delta }\left\vert 
\begin{array}{cc}
F_{\unit{u}u} & e_{lu} \\ 
F_{uv} & e_{lv}%
\end{array}%
\right\vert ,\quad l=1,...,\mu -1.
\end{equation}%
From equations (42) it follows that

\begin{equation}
\frac{\partial e_{l}(u,v)}{\partial t_{m}}-\frac{\partial e_{m}(u,v)}{%
\partial t_{l}}%
=e_{lu}u_{t_{m}}+e_{lv}v_{t_{m}}-e_{mu}u_{t_{l}}-e_{mv}v_{t_{l}}=0.
\end{equation}%
Hence, there exists a function $\varphi (t)$ such that

\begin{equation}
e_{l}(u,v)=\varphi _{t_{l}},\quad l=1,...,\mu -1.
\end{equation}

For D and E cases one has $e_{0}=1,e_{1}=x,e_{2}=y$. Thus, $\varphi
_{t_{0}}=1,u=\varphi _{t_{1}},v=\varphi _{t_{2}}$ and equations (45) are of
the form

\begin{equation}
\varphi _{t_{l}}=e_{l}(\varphi _{t_{1}},\varphi _{t_{2}}),\quad l=3,...,\mu
-1.
\end{equation}%
In terms of u and v one has the systems

\begin{equation}
u_{t_{l}}=\frac{\partial e_{l}(u,v)}{\partial t_{1}},\quad v_{t_{l}}=\frac{%
\partial e_{l}(u,v)}{\partial t_{2}},\quad l=3,...,\mu -1.
\end{equation}%
In virtue of (44) the systems (46),(47) describe commuting flows. From the
above equations one also concludes that for all D and E cases

\begin{equation}
\delta F(u,v)\div F(u,v;dt)=d\varphi .
\end{equation}%
The relations of the type (48) have appeared earlier within the study of
semiversal unfoldings of hypersurface singularities ( see e.g. \cite{Hert},
Chapter 5).

The systems (46) or (47) governs motion of critical points for versal
deformations of D and E singularities.

\ Concretely, for the $D_{n}$ case one has the system

\begin{equation}
\varphi _{t_{k}}=(\varphi _{t_{2}})^{k-1},\quad k=3,4,...,n-1
\end{equation}%
and $u=\varphi _{t_{1}},v=\varphi _{t_{2}}$. Effectively, it is again the
family of the BH and higher BH equations as in $A_{n}$ case.

For $E_{6}$ case one has the equations

\begin{equation}
\varphi _{t_{3}}=(\varphi _{t_{2}})^{2},\quad \varphi _{t_{4}}=\varphi
_{t_{1}}\varphi _{t_{2}},\quad \varphi _{t_{5}}=\varphi _{t_{1}}(\varphi
_{t_{2}})^{2}
\end{equation}%
or three systems

\begin{equation}
\begin{array}{c}
u_{t_{3}}=(v^{2})_{t_{1,}} \\ 
v_{t_{3}}=(v^{2})_{t_{2,}}%
\end{array}%
;\quad 
\begin{array}{c}
u_{t_{4}}=(uv)_{t_{1,}} \\ 
v_{t_{4}}=(uv)_{t_{2,}}%
\end{array}%
;\quad 
\begin{array}{c}
u_{t_{5}}=(uv^{2})_{t_{1,}} \\ 
v_{t_{5}}=(uv^{2})_{t_{2,}}%
\end{array}%
.
\end{equation}

For $E_{7}$ singularity (35) one has

\begin{equation}
\varphi _{t_{3}}=(\varphi _{t_{2}})^{2},\quad \varphi _{t_{4}}=(\varphi
_{t_{2}})^{3},\quad \varphi _{t_{5}}=(\varphi _{t_{2}})^{4},\quad \varphi
_{t_{6}}=\varphi _{t_{1}}\varphi _{t_{2}}\quad
\end{equation}%
or equivalently

\begin{equation}
\begin{array}{c}
u_{t_{3}}=(v^{2})_{t_{1,}} \\ 
v_{t_{3}}=(v^{2})_{t_{2,}}%
\end{array}%
;\quad 
\begin{array}{c}
u_{t_{4}}=(v^{3})_{t_{1,}} \\ 
v_{t_{4}}=(v^{3})_{t_{2,}}%
\end{array}%
;\quad 
\begin{array}{c}
u_{t_{5}}=(v^{4})_{t_{1,}} \\ 
v_{t_{5}}=(v^{4})_{t_{2,}}%
\end{array}%
;\quad 
\begin{array}{c}
u_{t_{6}}=(uv)_{t_{1,}} \\ 
v_{t_{6}}=(uv)_{t_{2,}}%
\end{array}%
.
\end{equation}

Finally, for $E_{8}$ singularity one has the system \ of five equations

\begin{equation}
\varphi _{t_{3}}=(\varphi _{t_{2}})^{2},\quad \varphi _{t_{4}}=(\varphi
_{t_{2}})^{3},\quad \varphi _{t_{5}}=\varphi _{t_{1}}\varphi _{t_{2}},\quad
\varphi _{t_{6}}=\varphi _{t_{1}}(\varphi _{t_{2}})^{2},\quad \varphi
_{t_{7}}=\varphi _{t_{1}}(\varphi _{t_{2}})^{3}
\end{equation}%
or the systems

\begin{equation}
\begin{array}{c}
u_{t_{3}}=(v^{2})_{t_{1,}} \\ 
v_{t_{3}}=(v^{2})_{t_{2,}}%
\end{array}%
;%
\begin{array}{c}
u_{t_{4}}=(v^{3})_{t_{1,}} \\ 
v_{t_{4}}=(v^{3})_{t_{2,}}%
\end{array}%
;%
\begin{array}{c}
u_{t_{5}}=(uv)_{t_{1,}} \\ 
v_{t_{5}}=(uv)_{t_{2,}}%
\end{array}%
;%
\begin{array}{c}
u_{t_{6}}=(uv^{2})_{t_{1,}} \\ 
v_{t_{6}}=(uv^{2})_{t_{2,}}%
\end{array}%
;%
\begin{array}{c}
u_{t_{7}}=(uv^{3})_{t_{1,}} \\ 
v_{t_{7}}=(uv^{3})_{t_{2,}}%
\end{array}%
.
\end{equation}%
Interrelations between the systems (50)-(55) and their 1+1-dimensional
reductions are of interest. It would be also of interest to analyse a
connection between apparently different systems of equations corresponding
to different choices of a basis for the local algebra $Q_{F_{0}}$ of the
critical points.

Analysis presented above can be easily extended to the \ singularities and
their versal unfoldings of any corank m. One derives that the set of
critical points $u_{i}(i=1,...,m)$ are components of the gradient $%
u_{i}=\varphi _{t_{i}}(i=1,...,m)$ and the function $\varphi
(t_{1},...,t_{\mu -1})$ obeys the system of equations

\begin{equation}
\varphi _{t_{k}}=e_{k}(\varphi _{t_{1}},...,\varphi _{t_{m}}),\quad
k=m+1,...,\mu -1
\end{equation}%
where $e_{k}(x_{1},...,x_{m}),k=0,1,...,\mu -1$ form a basis of the local
algebra $Q_{F_{0}}$ for the critical point of the function $%
F_{0}(x_{1},...,x_{n})$. Equivalently, one has the systems

\begin{equation}
\frac{\partial u_{i}}{\partial t_{k}}=\frac{\partial e_{k}(u_{1},...,u_{m})}{%
\partial t_{i}},\quad i=1,...,m;\quad k=m+1,...,\mu -1
\end{equation}%
and also

\begin{equation}
F(u_{1},...,u_{m};dt)=d\varphi .
\end{equation}

Properties of these equations will be discussed elsewhere.

\section{Nonversal unfoldings and integrable systems of hydrodynamic type.}

\bigskip

Construction given in the previous section is applicable to nonversal
unfoldings too. For nonversal \ unfoldings, even if the infinitesimal
deformation is the form (39), the functions $e_{k}$ may not form a basis of
the local algebra or number of parameters of deformation may be less than $%
\mu $. Nevertheless, it is not difficult to show that for such unfoldings of
corank two singularities the general formulas (40)-(45) remain unaltered,
only number of the variables $t_{k}$ can be different.

Here we will consider a particular unfolding of the umbilic singularities
given by the function

\begin{equation}
F(x,y;t_{1},t_{3})=\alpha x^{3}+3xy^{2}+t_{3}(\beta x^{2}+\gamma
y^{2})+t_{1}x
\end{equation}%
where $\alpha ,\beta ,\gamma $ are parameters. At $\alpha =\pm 1$ \ and $%
\beta \pm \gamma =0$ this unfolding is equivalent to the versal unfoldings
of $D^{\pm }$ singularity with the "frozen" parameters $t_{2}=0$ and $%
t_{0}=0.$ At $\alpha =0$ the function (59) represents the two-dimensional
part of the infinite dimensional deformation of the germ $3xy^{2}$ of the
critical point with infinite multiplicity \cite{Arnold2}.

We will be interested in the subclass of unfoldings (59) for which $%
e_{3}=\beta x^{2}+\gamma y^{2}$ belongs to the tangent space of the germ $%
\alpha x^{3}+3xy^{2}$, i.e. $e_{3}\sim F_{0x}$. This contraint is verified
if $\alpha \gamma =\beta $. \ In this case the function F after trivial
rescaling takes the form

\begin{equation}
F(x,y;t_{1},t_{3})=\alpha x^{3}+3xy^{2}+t_{3}(\alpha x^{2}+y^{2})+t_{1}x.
\end{equation}%
Repeating the calculation performed in the previous section, one gets the
system

\begin{equation}
u_{t_{1}}=-\frac{F_{vv}}{\Delta },\quad v_{t_{1}}=\frac{F_{uv}}{\Delta },
\end{equation}

\begin{equation}
\begin{array}{c}
u_{t_{3}}=-2\alpha u\frac{F_{vv}}{\Delta }+2v\frac{F_{uv}}{\Delta },\medskip
\\ 
v_{t_{3}}=-2v\frac{F_{\unit{u}u}}{\Delta }+2\alpha u\frac{F_{uv}}{\Delta }.%
\end{array}%
\end{equation}%
Using (61), one can rewrite the last system as

\begin{equation}
\begin{array}{c}
u_{t_{3}}=(\alpha u^{2}+v^{2})_{t_{1}},\medskip \\ 
v_{t_{3}}=-2v\frac{F_{\unit{u}u}}{\Delta }+2\alpha uv_{t_{1}}.%
\end{array}%
\end{equation}%
This system is not of hydrodynamical type. In absence of $t_{2}$ equations
(18) we cannot express $\frac{F_{\unit{u}u}}{\Delta }$ in terms of u and v
and their derivaties. A way to get the hydrodynamical type system is to
impose a constraint on F. It is not difficult to show that among the
constraints having form of linear relation between second order derivatives
of F only the constraint $F_{\unit{u}u}=\delta F_{vv}$ where $\delta $ is a
constant is admissable ( nontrivial). Under this constraint $\frac{F_{\unit{u%
}u}}{\Delta }=\delta \frac{F_{vv}}{\Delta }=-\delta u_{t_{1}}$ and the
system (63) becomes

\begin{equation}
\left( 
\begin{array}{c}
u \\ 
v%
\end{array}%
\right) _{t_{3}}=2\left( 
\begin{array}{cc}
\alpha u, & v \\ 
\delta v, & \alpha u%
\end{array}%
\right) \left( 
\begin{array}{c}
u \\ 
v%
\end{array}%
\right) _{t_{1}}.
\end{equation}%
The first equation of this system, i.e. $u_{t_{3}}=(\alpha
u^{2}+v^{2})_{t_{1}}$ implies the existence of a function $\varphi
(t_{1},t_{3})$ such that $u=\varphi _{t_{1}}$ and $\alpha
u^{2}+v^{2}=\varphi _{t_{3}}$. As the consequence one has

\begin{equation*}
F(u,v;dt)=F(u,v;t+dt)-F(u,v;t)=t_{3}(\alpha u^{2}+v^{2})+t_{1}u=d\varphi .
\end{equation*}

Among two parameters in this system only one is relevant. Indeed for
nonvanishing $\delta $ the rescalling $u\rightarrow u,v^{2}\rightarrow
\left\vert \delta \right\vert v^{2},t_{1}\rightarrow \left\vert \delta
\right\vert t_{1},\alpha \rightarrow \frac{\alpha }{\left\vert \delta
\right\vert }$ converts the system (64) into the same system with $\delta
=\pm 1$ leaving only the parameter $\alpha $ free. For nonvanishing $\alpha $
one can convert the system (64) into that with $\alpha =1$ and free $\delta $%
. \ Since we will consider the case $\alpha =0$ among the others the first
choice is preferable.

The system (64) with $\delta =\pm 1$ and particular values of $\alpha $
coincides with some well-known integrable hydrodynamic type systems. Indeed,
at $\alpha =0$ one has the system

\begin{equation}
u_{t_{3}}=2vv_{t_{1}},\quad v_{t_{3}}=2\delta vu_{t_{1}}
\end{equation}%
which is equivalent to the dispersionless Toda equation $\Phi
_{t_{3}t_{3}}=4\delta (\exp \Phi )_{t_{1}t_{1}}$ for $\Phi =\log v^{2}$. \
For $\alpha =2$ and $\delta =1$ it is the hyperbolic one-layer Benney system
( classical shallow water equation) for variables $u$ and $v^{2}$ ( see e.g.%
\cite{Whitham} and \cite{Zakharov}). At $\alpha =2$ and $\delta =-1$ it is
the elliptic one-layer Benney system which is equivalent to the
dispersionless Da Rios system. For $\alpha =\delta =\pm 1$ the corresponding
system $u_{t_{3}}=(\pm u^{2}+v^{2})_{t_{1}},v_{t_{3}}=\pm 2(uv)_{t_{1}}$
decomposes into two BH equations $(u\pm \sqrt{\pm }v)_{t_{3}}=\mp ((u\pm 
\sqrt{\pm }v)^{2})_{t_{1}}$. For other values of $\alpha $ the system (64)
is equivalent to the system ( $t=-2\alpha t_{3},x=t_{1},\rho =v^{\alpha
\delta })$

\begin{equation}
\rho _{t}+(\rho u)_{x}=0,\quad u_{t}+u_{x}u+\frac{p_{x}}{\rho }=0,
\end{equation}%
which describes one-dimensional motions of ideal barotropic gas with the
density $\rho $ and pressure $p=\frac{1}{\alpha (\alpha \delta +2)}\rho ^{%
\frac{\alpha \delta +2}{\alpha \delta }}$ (see e.g. \cite{Rozd}).

In terms of Riemann invariants $\beta _{1}=u+\frac{1}{\sqrt{\delta }}v,\beta
_{2}=u-\frac{1}{\sqrt{\delta }}v$ the system (64) is of the form

\begin{equation}
\beta _{1t_{3}}=\left\vert \delta \right\vert (\varepsilon (\beta _{1}+\beta
_{2})+\beta _{1})\beta _{1t_{1}},\quad \beta _{2t_{3}}=\left\vert \delta
\right\vert (\varepsilon (\beta _{1}+\beta _{2})+\beta _{2})\beta _{2t_{1}},
\end{equation}%
where $\varepsilon =\frac{1}{2}(\frac{\alpha }{\left\vert \delta \right\vert 
}-1)$. This form of the system (64) clearly shows that only the ratio $\frac{%
\alpha }{\left\vert \delta \right\vert }$ is relevant. We will put \ $\delta
=1$ in what follows. The hodograph equations are of the form $F_{\beta
_{1}}=0,F_{\beta _{2}}=0$ while the constraint $F_{\unit{u}u}=F_{vv}$
becomes $F_{\beta _{1}\beta _{2}}=0$. Under this condition $\Delta =F_{\beta
_{1}\beta _{1}}F_{\beta _{2}\beta _{2}}$ and hence the catastrophe set where
the solutions of (64) or (67) exhibit gradient catastrophe decomposes into
two pieces $F_{\beta _{1}\beta _{1}}=0$ and $F_{\beta _{2}\beta _{2}}=0$.

The system (67) represents a two-component case of the so-called $%
\varepsilon $-systems discussed in \cite{Pavl} . In a different manner the
systems (64)-(66) arozen within the study of hydrodynamic type systems
associated with the two dimensional Frobenius manifolds \cite{Dubr1}.

\section{Hodograph equations for hydrodynamic type systems as equations for
critical points and Euler-Poisson-Darboux equation}

\ \ \ \ \ \ \ \ \ \ \ \ \ \ \ \ \ \ \ \ \ \ \ \ \ \ \ \ \ \ \ \ \ \ \ \ \ \
\ \ \ \ \ \ \ \ \ \ \ \ \ \ \ \ \ \ \ \ \ \ \ \ \ \ \ \ \ \ \ \ \ \ \ \ \ \
\ \ \ \ \ \ \ \ \ \ \ \ \ \ \ \ \ \ \ \ \ \ \ \ \ \ \ \ \ \ \ \ \ \ \ \ \ \
\ \ \ \ \ \ \ \ \ \ \ \ \ \ \ \ \ \ \ \ \ \ \ \ \ \ \ \ \ \ \ \ \ \ \ \ \ \
\ \ \ \ \ \ \ \ \ \ \ \ \ \ \ \ \ \ \ \ \ \ \ \ \ \ \ \ \ \ \ \ \ \ \ \ \ \
\ \ \ \ \ \ \ \ \ \ \ \ \ \ \ \ \ \ \ \ \ \ \ \ \ \ \ \ \ \ \ \ \ \ \ \ \ \
\ \ \ \ \ \ \ \ \ \ \ \ \ \ \ \ \ \ \ \ \ \ \ \ \ \ \ \ \ \ \ \ \ \ \ \ \ \
\ \ \ \ \ \ \ \ \ \ \ \ \ \ \ \ \ \ \ \ \ \ \ \ \ \ \ \ \ \ \ \ \ \ \ \ \ \
\ \ \ \ \ \ \ \ \ \ \ \ \ \ \ \ \ \ \ \ \ \ \ \ \ \ \ \ \ \ \ \ \ \ \ \ \ \
\ \ \ \ \ \ \ \ \ \ \ \ \ \ 

\bigskip Critical points for unfoldings of A and D singularities as the
functions of deformation parameters represent very particular solutions of
the BH equation and the system (64). What about the other solutions of these
integrable equations? Do they describe certain unfoldings of critical points
for other singularities?

To address this question we begin with the BH equation $%
u_{t_{2}}=(u^{2})_{t_{1}}$. A standard hodograph equation for it (see e.g. 
\cite{Whitham, Rozd})

\begin{equation}
t_{1}+2t_{2}u+f(u)=0
\end{equation}%
is the equation defining critical point $u(t)$ of the function $%
W(x;t)=F_{0}(x)+t_{2}x^{2}+t_{1}x$ where $F_{0x}(x)=f(x)$ and $f(u)$ is the
function inverse to the initial data function $u_{0}(t_{1})=u(t_{1},t_{2}=0)$%
. Let us consider a family of BH solutions which corresponds to the family
of initial data with $%
F_{0}(x)=x^{n+1}+t_{n}x^{n}+t_{n-1}x^{n-1}+...+t_{3}x^{3}$ where $%
t_{3},...,t_{n}$ are parameters and n is arbitrary. These solutions $%
u(t_{1,}t_{2},t_{3},...,t_{n})$ of BH equation \ clearly are associated with
unfoldings ( versal at $t_{n}=0$) of $A_{n}$ singularities considered in the
previous section. \ Thus, the class of solutions of BH equation which
corresponds to the family of initial data of this type with unbounded n
describes unfoldings of critical points for all $A_{n}$ singularities.
Alternatively, the parameters $t_{3},...,t_{n}$ can be viewed as times for
higher BH equations within the infinite BH hierarchy ($n\rightarrow \infty $%
) \cite{Kodama-Kon}.

An extension of this construction to the system (64) is almost
straightforward. We already saw that the particular solution of this system
describes motion of the critical points for the function (60). Thus, we will
look for the functions $F(x,y;t)=F_{0}(x,y)+$ $t_{3}(\alpha
x^{2}+y^{2})+t_{1}x$ with $F_{0}$ different \ from $\alpha x^{3}+3xy^{2}$. \
It is convenient to pass to the variables $x_{1}=x+y,$ $x_{2}=x-y$. In these
variables $F=F_{0}(x_{1},x_{2})+\frac{t_{2}}{4}((\varepsilon
+1)x_{1}^{2}+2\varepsilon x_{1}x_{2}+(\varepsilon +1)x_{2}^{2})+\frac{t_{1}}{%
2}(x_{1}+x_{2})$ where $\varepsilon =\frac{1}{2}(\alpha -1)$ and $%
t_{2}=2t_{3}$. Equations defining the critical points $\beta _{1}$and $\beta
_{2}$ are

\begin{equation}
F_{\beta _{1}}=0,\quad F_{\beta _{2}}=0.
\end{equation}

\textbf{Lemma}. Solution of equations (69), with the function F of the form
given above and any function $F_{0}$ such that $F_{\beta _{1}\beta _{2}}=0$
and $F_{\beta _{1}\beta _{1}}F_{\beta _{2}\beta _{2}}\neq 0$, is a solution
of the system (67).

Proof. Differentiating equation (69) w.r.t. $t_{1}$ and $t_{2}$ and assuming
that $F_{\beta _{1}\beta _{2}}=0$, one gets

\begin{equation}
\beta _{1t_{1}}=-\frac{1}{2F_{\beta _{1}\beta _{1}}},\quad \beta _{2t_{1}}=-%
\frac{1}{2F_{\beta _{2}\beta _{2}}}
\end{equation}%
and

\begin{equation}
\beta _{1t_{2}}=-\frac{((\varepsilon +1)\beta _{1}+\varepsilon \beta _{2})}{%
2F_{\beta _{1}\beta _{1}}},\quad \beta _{2t_{2}}=-\frac{(\varepsilon \beta
_{1}+(\varepsilon +1)\beta _{2})}{2F_{\beta _{2}\beta _{2}}}.
\end{equation}%
Eliminating $F_{\beta _{1}\beta _{1}}$ and $F_{\beta _{2}\beta _{2}}$ from
equations (71), one obtains the system (67).$\square $

\ \ The function $F_{0}$ due to the relations

\begin{equation*}
\beta _{1t_{1}}(t_{1},t_{2}=0)=-\frac{1}{2F_{0\beta _{1}\beta
_{1}}|_{t_{2}=0}},\quad \beta _{2t_{1}}(t_{1},t_{2}=0)=-\frac{1}{2F_{0\beta
_{2}\beta _{2}}|_{t_{2}=0}}
\end{equation*}%
is defined implicitly by the initial data for $\beta _{1}$ and $\beta _{2}$.
The functions F and aF where a is an arbitrary constant, obviously, give
rise to the same equation.

Equations (67) imply that

\begin{equation*}
\frac{1}{2}(\beta _{1}+\beta _{2})_{t_{2}}=\frac{1}{4}\left( (\varepsilon
+1)\beta _{1}^{2}+2\varepsilon \beta _{1}\beta _{2}+(\varepsilon +1)\beta
_{2}^{2}\right) _{t_{1}}.
\end{equation*}%
So, $\frac{1}{2}(\beta _{1}+\beta _{2})=\varphi _{t_{1}},\frac{1}{4}\left(
(\varepsilon +1)\beta _{1}^{2}+2\varepsilon \beta _{1}\beta
_{2}+(\varepsilon +1)\beta _{2}^{2}\right) =\varphi _{t_{2}}$ where $\varphi 
$ is a function of $t_{1}$ and $t_{2}$ and hence

\begin{equation}
F(\beta _{1},\beta _{2};dt)=d\varphi .
\end{equation}

There is a special subclass of functions F of particular interest. It is
given by

\begin{equation}
W(x_{1},x_{2};t)=\frac{1}{2\varepsilon }\doint\limits_{\gamma }\frac{%
d\lambda }{2\pi i}\left( \sum_{k\geq 1}\lambda ^{k-1}t_{k}\right) \left( (1-%
\frac{x_{1}}{\lambda })(1-\frac{x_{2}}{\lambda })\right) ^{-\varepsilon }
\end{equation}%
where $\gamma $ denotes a large positively oriented circle on the $\lambda $
plane. \ Written explicitly the function W is the series

\bigskip

\begin{eqnarray}
W &=&\frac{1}{2}t_{1}(x_{1}+x_{2})+t_{2}\frac{1}{4}\left[ (\varepsilon
+1)(x_{1}^{2}+x_{2}^{2})+2\varepsilon x_{1}x_{2}\right] +  \notag \\
&&+t_{3}\frac{1}{12}(\varepsilon +1)\left[ (\varepsilon
+2)(x_{1}^{3}+x_{2}^{3})+3\varepsilon (x_{1}x_{2}^{2}+x_{2}x_{1}^{2})\right]
+ \\
&&+t_{4}\frac{1}{48}(\varepsilon +1)\left[ (\varepsilon +2)(\varepsilon
+3)(x_{1}^{4}+x_{2}^{4})+4\varepsilon (\varepsilon
+2)(x_{1}x_{2}^{3}+x_{2}x_{1}^{3})+6\varepsilon (\varepsilon
+1)x_{1}^{2}x_{2}^{2}\right] +...  \notag
\end{eqnarray}%
An important property of the function W is that it obeys the
Euler-Poisson-Darboux equation $E(\varepsilon ,\varepsilon )$, i.e.

\begin{equation}
(x_{1}-x_{2})W_{x_{1}x_{2}}=\varepsilon (W_{x_{1}}-W_{x_{2}}).
\end{equation}%
This equation and representation of its solutions in the form (73) are known
for more than a century (see \cite{Darboux}). In the papers \cite%
{Kon-Alonso-Med1,Kon-Alonso-Med2,Kon-Alonso-Med3} it was observed that the
hodograph equations for the one-layer Benney hierarchy ($\varepsilon =\frac{1%
}{2}$) and dispersionless dToda hierarchy ($\varepsilon =-\frac{1}{2}$) \
are nothing but that the equations for critical points of the function $%
2\varepsilon W$.

For arbitrary $\varepsilon \neq 0$ equations

\begin{equation}
W_{\beta _{1}}=0,\quad W_{\beta _{2}}=0
\end{equation}%
for the critical points are the hodograph equations for the system (67). \
Due to equation (75) the function W at $\beta _{1}\neq \beta _{2}$
automatically verifies the condition $W_{\beta _{1}\beta _{2}}=0$. \ For the
function F of the form (73) the function $F_{0}$ is a special one. The
variables $t_{3},t_{4},...$ can be viewed as the variables parametrizing a
family of initial data for the system (67). Alternatively, one can treat
them as the higher "times" for the commuting systems

\begin{equation}
\beta _{1t_{k}}=\theta _{1k}(\beta _{1},\beta _{2})\beta _{1t_{1}},\quad
\beta _{2t_{k}}=\theta _{2k}(\beta _{1},\beta _{2})\beta _{2t_{1}},\quad
k=3,4,...
\end{equation}%
with the characteristic velocities $\theta _{1k}(\beta _{1},\beta _{2})=%
\frac{\partial }{\partial x_{1}}(\frac{\partial W}{\partial t_{k}}%
)|_{x=\beta },\quad \theta _{2k}(\beta _{1},\beta _{2})=\frac{\partial }{%
\partial x_{2}}(\frac{\partial W}{\partial t_{k}})|_{x=\beta }$. \ The
totality of these systems is an infinite hierarchy of systems associated
with the system (67) and the function W plays the role of the generating
function for this hierarchy.

In the singular case $\varepsilon =0$ \ the generating function W$%
_{\varepsilon =0}$ is given by

\begin{equation}
W_{\varepsilon =0}=\doint\limits_{\gamma }\frac{d\lambda }{2\pi i}\left(
\sum_{k\geq 1}\lambda ^{k-1}t_{k}\right) \log \left( (1-\frac{x_{1}}{\lambda 
})(1-\frac{x_{2}}{\lambda })\right) =-\sum_{n\geq 1}\frac{1}{n}%
t_{n}(x_{1}^{n}+x_{2}^{n}).
\end{equation}%
which gives rise to two independent BH hierarchies for $\beta _{1}$ and $%
\beta _{2}$.

The observation that the hodograph solutions of the two-component
hydrodynamic systems of $\varepsilon $-type describe critical points of
function (73) is extendable to the multi-component case. It was shown in 
\cite{Kon-Alonso-Med1} that the critical points of the function

\begin{equation}
W(x_{1},x_{2},...,x_{n};t)=\frac{1}{2\varepsilon }\doint\limits_{\gamma }%
\frac{d\lambda }{2\pi i}\left( \sum_{k\geq 1}\lambda ^{k-1}t_{k}\right)
\left( (1-\frac{x_{1}}{\lambda })(1-\frac{x_{2}}{\lambda })\cdot \cdot \cdot
(1-\frac{x_{n}}{\lambda })\right) ^{-\varepsilon }
\end{equation}%
at $\varepsilon =\frac{1}{2}$ are described by hodograph solutions of the
dispersionless n-component coupled KdV equation.

For arbitrary $\varepsilon \neq 0$ the function (79) obeys the
Euler-Poisson-Darboux system

\begin{equation}
(x_{i}-x_{k})W_{x_{i}x_{k}}=\varepsilon (W_{x_{i}}-W_{x_{k}}),\quad i\neq
k;i,k,=1,...,n.
\end{equation}%
From the equations $F_{\beta _{i}}=0,i=1,...,n$ for critical points $\beta
_{1},...,\beta _{n}$ of the function (79) one finds $\beta _{it_{k}}=-\frac{%
W_{k\beta _{i}}}{W_{\beta _{i}\beta _{i}}},i=1,...,n;k=1,2,...$ where $W_{k}$
are defined by the expansion $W(x;t)=\sum_{k\geq 1}t_{k}W_{k}(x)$. Since $%
W_{1\beta _{i}}=1$ one gets the following hierarchy of hydrodynamic type
systems governing the motion of the critical points

\begin{equation}
\beta _{it_{k}}=W_{k\beta _{i}}(\beta )\beta _{it_{1}},\quad i=1,...,n;\quad
k=2,3,...
\end{equation}%
The first member of this hierarchy is given by the $\varepsilon $-system $%
\beta _{it_{2}}=(\varepsilon (\sum_{m=1}^{m=n}\beta _{m})+\beta _{i})\beta
_{it_{1}},\quad i=1,...,n$ considered in \cite{Pavl}. The system (81)
implies that $W_{kt_{l}}-W_{lt_{k}}=0,\quad k,l=1,2,...$. So, $W_{k}=\varphi
_{t_{k}}$ and hence

\begin{equation*}
W(\beta ;dt)=d\varphi .
\end{equation*}%
It was observed in \cite{Kon-Alonso-Med2} ( see also \cite{Pavl}) that the
systems (81) have an interesting property: densities $P$ of their conserved
quantities satisfy the Euler-Poisson-Darboux equations dual to (80) ( i.e.
with opposite sign of $\varepsilon $). An infinite family of such densities
is given by

\begin{equation*}
P_{n}(\beta )=\doint\limits_{\gamma }\frac{d\lambda }{2\pi i}\lambda
^{n}\left( (1-\frac{\beta _{1}}{\lambda })(1-\frac{\beta _{2}}{\lambda }%
)\cdot \cdot \cdot (1-\frac{\beta _{n}}{\lambda })\right) ^{\varepsilon
},\quad n=0,1,2,...
\end{equation*}

A class of the hydrodynamic type systems for which hodograph equations
coincide with the equations for critical points of certain functions is, in
fact, larger. It consists of all semihamiltonian diagonal systems for which
characteristic velocities $\theta _{l}$ are components of the gradient of a
function, i.e.

\begin{equation}
\beta _{lt}=\Phi _{\beta _{l}}\beta _{lx},\quad l=1,...,n.
\end{equation}%
where $\Phi _{\beta _{l}}\div \frac{\partial \Phi }{\partial \beta _{l}}$.
Indeed, according to \cite{Tsarev} the generalized hodograph equations for
semihamiltonian diagonal system $\beta _{lt}=\theta _{l}(\beta )\beta
_{lx},l=1,...,n$ are given by the system

\begin{equation}
x+\theta _{l}(\beta )t+\omega _{l}(\beta )=0,\quad l=1,...,n
\end{equation}%
where the functions $\omega _{l}(\beta )$ obeys the equations

\begin{equation}
\frac{\omega _{i\beta _{k}}}{\omega _{k}-\omega _{i}}=\frac{\theta _{i\beta
_{k}}}{\theta _{k}-\theta _{i}},\quad i\neq k
\end{equation}%
If $\theta _{i}=\Phi _{\beta _{i}}$then the l.h.s. of (84) is skew symmetric
and hence $\omega _{i\beta _{k}}=\omega _{k\beta _{i}}$. So there exists a
function $\Phi _{\omega }$ such that $\omega _{i}(\beta )=\Phi _{\omega
\beta _{i}},i=1,..,n$. \ As a result, the hodograph equations (83) take the
form

\begin{equation*}
x+\Phi _{\beta _{l}}t+\Phi _{\omega \beta _{l}}=0,\quad l=1,...,n
\end{equation*}%
that coincides with the equations $F_{\beta _{l}}=0$ for critical points of
the function

\begin{equation}
W=x(x_{1}+...x_{n})+t\Phi (x)+\Phi _{\omega }(x).
\end{equation}

From (84) it follows that this function obeys the system of equations

\begin{equation}
W_{x_{i}x_{k}}=\frac{\Phi _{x_{i}x_{k}}}{\Phi _{x_{i}}-\Phi _{x_{k}}}%
(W_{x_{i}}-W_{x_{k}}),\quad i\neq k;\quad i,k,=1,...,n.
\end{equation}%
as well as the function $\Phi _{\omega }$. Densities $P$ of the conserved
quantities for the system (82) satisfy the equations \cite{Tsarev}

\begin{equation}
P_{\beta _{i}\beta _{k}}=-\frac{\Phi _{\beta _{i}\beta _{k}}}{\Phi _{\beta
_{i}}-\Phi _{\beta _{k}}}(P_{\beta _{i}}-P_{\beta _{k}}),\quad i\neq k;\quad
i,k,=1,...,n.
\end{equation}
Thus, the hydrodynamic type system (82) describes critical points of the
function W (85) which obeys equations (86) and this equations (as well as
the equations for the generating function $\Phi _{\omega }$ of symmetries)
and the equations for conserved densities P of (82) are dual to each other.

\section{Hierarchies of integrable systems and normal forms of singularities}

\bigskip

Now let us discuss the hierarchies of two- and three-component integrable
systems considered in the previous section and the corresponding functions W
from the singularity theory viewpoint. The functions $W=\sum_{k\geq
1}t_{k}W_{k}$ provide us with an infinite families of symmetric homogeneous
functions $W_{k}(x_{1},...,x_{n})$ of degrees k. Is there any relation
between these functions and normal forms of germs in singularity theory? We
will present here two observations which indicate that such a connection
exists.

In the two-component case the function W (73) in the variables x and y is
given by

\begin{equation}
W(x,y;t)=\frac{1}{2\varepsilon }\doint\limits_{\gamma }\frac{d\lambda }{2\pi
i}\left( \sum_{k\geq 1}\lambda ^{k-1}t_{k}\right) \left( (1-2x\frac{1}{%
\lambda }+(x^{2}-y^{2})\frac{1}{\lambda ^{2}}\right) ^{-\varepsilon }
\end{equation}%
or

\begin{equation}
W=t_{1}x+\frac{1}{2}t_{2}(\alpha x^{2}+y^{2})+\frac{1}{6}(\alpha
+1)t_{3}(\alpha x^{3}+3xy^{2})+\frac{1}{384}(\alpha
+1)t_{4}(ax^{4}+bx^{2}y^{2}+cy^{4})+...
\end{equation}%
where $a=15\alpha ^{2}+24\alpha -15,\quad b=\alpha ^{2}+8\alpha -33,\quad
c=6\alpha ^{2}-48\alpha -102$.

Third term in (89) generates the first higher commuting flow for the system
(64). At the same time, it provides us with the normal form of simple $D_{4}$
singularity after the trivial rescaling of x. Fourth term gives rise to the
next commuting flow for (64). Does it correspond to some standard normal
form? It is easy to see that, rescaling x and y, one can convert it into the
form $x^{4}+dx^{2}y^{2}+y^{4}$ where d is certain function of $\alpha $ .
For generic $\alpha $ the parameter $d\neq \pm 2$. Thus one has the normal
form of the unimodular $X_{9}$ singularity (see \cite{Arnold4,AGV1}). Higher 
$W_{k}$ in the expansion $W=\sum_{k\geq 1}t_{k}W_{k}$ which generate higher
flows are of the form $W_{k}=\sum_{l=0}^{l=\left[ \frac{k}{2}\right]
}a_{l}x^{k-2l}y^{2l}$ where $a_{l}(\alpha )$ are certain polynomials in $%
\alpha .$ So, they describe unimodular singularities of the order k with the
germs symmetric with respect to the reflection $y\rightarrow -y$.

Second example is given by the system (81) with $n=3$ and $\varepsilon =%
\frac{1}{2}$, i.e. by the dispersionless three-component coupled KdV system.
The function W is

\begin{eqnarray*}
W &=&\frac{1}{2}t_{1}(x_{1}+x_{2}+x_{3})+\frac{1}{8}%
t_{2}(3x_{1}^{2}+3x_{2}^{2}+3x_{3}^{2}+2x_{1}x_{2}+2x_{1}x_{3}+2x_{2}x_{3})+
\\
&&+\frac{1}{16}%
t_{3}(5x_{1}^{3}+5x_{2}^{3}+5x_{3}^{3}+3x_{1}^{2}x_{2}+3x_{1}^{2}x_{3}+3x_{2}^{2}x_{1}+3x_{2}^{2}x_{3}+3x_{3}^{2}x_{1}+3x_{3}^{2}x_{2}+2x_{1}x_{2}x_{3})+...
\end{eqnarray*}%
In terms of variables x,y,z defined by

\begin{equation}
x_{1}=ax+y+z,\quad x_{2}=ax+qy+q^{2}z,\quad x_{3}=ax+q^{2}y+qz,
\end{equation}%
where q=$\exp (\frac{2\pi i}{3})$ and $a=\frac{2}{\sqrt[3]{35}},$ \ one has

\begin{equation}
W=\frac{3}{2}at_{1}x+\frac{3}{8}t_{2}(5a^{2}x^{2}+4yz)+\frac{1}{2^{5}3^{3}}%
t_{3}(x^{3}+y^{3}+z^{3}+\frac{21}{2}axyz)+...
\end{equation}%
The transformation (90) is, in fact, the well-known relation between roots $%
x_{i}$ of a cubic equation and its Lagrange resolvents $\ell _{i}$ modulo
the identification $ax=\frac{1}{3}\ell _{1},y=\frac{1}{3}\ell _{2},z=\frac{1%
}{3}\ell _{3}$ (see e.g.\cite{vander}).

The $W_{3}$ term in (91) represents the normal form of the unimodular
parabolic singularity $P_{8}$ ( or $T_{3,3,3}$ ) of corank three (see \cite%
{Arnold4,AGV1}). At $t_{k}=0,k=4,5,...$ the function (91) gives the
unfolding of $P_{8}$ singularity. The dependence of the critical points
u,v,w for this function on deformation parameters $t_{1}$ and $t_{2}$ is
described by the three-component dispersionless coupled KdV system (81) or
equivalently by the system

\begin{equation}
\left( 
\begin{array}{c}
u \\ 
v \\ 
w%
\end{array}%
\right) _{t_{2}}=\left( 
\begin{array}{ccc}
\frac{5}{2}au, & \frac{1}{a}w, & \frac{1}{a}v \\ 
av, & \frac{5}{2}u, & w \\ 
aw, & v, & \frac{5}{2}u%
\end{array}%
\right) \left( 
\begin{array}{c}
u \\ 
v \\ 
w%
\end{array}%
\right) _{t_{1}}.
\end{equation}

The functions $W_{k}$ with $k=4,5,...$ represent higher order singularities
of corank three and the function (91) give their unfoldings.

\bigskip

\section*{Acknowledgements}

Author thanks Y. Kodama, \ F. Magri, \ G. Ortenzi and M. Pavlov for the
useful discussions and help. This work was partialy supported by the PRIN
2008 grant No. 28002K9KXZ.


\addcontentsline{toc}{section}{References}

\bigskip

\begin{thebibliography}{99}
\bibitem{Thom} \emph{R. Thom,}\newblock Structural Stability and
Morphogenesis, Benjamin-Addison, New York, 1975.

\bibitem{Lu} \emph{Yung-Chen Lu,}\newblock Singularity Theory and an
Introduction to Catastrophe Theory, Springer-Verlag, New York, 1976.

\bibitem{poston} \emph{T. Poston, I. Stewart,}\newblock Catastrophe Theory
and its Applications, Pitman, London, 1978.

\bibitem{AGV1} \emph{V.I. Arnold, S. M. Gusein-Zade, A.N. Varchenko,}%
\newblock Singularities of differentiable maps, vol.1, Birkhauser, Boston,
1985.

\bibitem{AGV2} \emph{V.I. Arnold, S. M. Gusein-Zade, A.N. Varchenko,}%
\newblock Singularities of differentiable maps, vol.II, Birkhauser, Boston,
1988.

\bibitem{Zeeman} \emph{E.C. Zeeman,}\newblock Breaking of waves, in: Proc.
of Symposium on Diff. Equations and Dynamics, Univ. of Warwick, 1968;
Lecture Notes in Math., vol. \textbf{206}, pp. 2-6, Springer-Verlag, New
York, 1971.

\bibitem{Guck} \emph{J. Guckenheimer,}\newblock\ Catastrophes and partial
differential equations, Ann. Inst. Fourier. Grenoble, \textbf{23} (2), 31-59
(1973).

\bibitem{Arnold2} \emph{V.I. Arnold,}\newblock Normal forms for functions
near degenerate critical points, the Weyl groups of $A_{k},D_{k},E_{k}$ and
Lagragian singularities, Funct. Anal.Appl., \textbf{6}, 254-272, (1972).

\bibitem{Arnold3} \emph{V.I. Arnold,}\newblock\ Critical points of smooth
functions and their normal forms, Russian Math. Survey, \textbf{30}, (5),
1-75 (1975).

\bibitem{Janich} \emph{K.Janich,}\newblock Caustics and catastrophes, Math.
Ann., \textbf{209},161-180 (1974).

\bibitem{Berry} \emph{M.V. Berry,}\newblock Waves and Thom's theorem, Adv.
Phys., \textbf{25}, 1-26 (1976).

\bibitem{Givental1} \emph{A.Givental,}\newblock Manifolds of polynomials
having a root of fixed multiplicity and generalized Newton equation, Funct.
Anal.Appl., \textbf{16 }(1), 10-14 (1982).

\bibitem{Malikov} \emph{Kh.M. Malikov,}\newblock Over-determination of a
system of differential equations for versal integrals of type A,D,E,
Differential equations, \textbf{18} (8), 986-991 (1982).

\bibitem{Zakal} \emph{V.M. Zakalyukin,}\newblock Reconstruction of fronts
and caustics depending on a parameter and versality of mappings, Journal of
Sov.Math., \textbf{27} (3), 2713-2735 (1984).

\bibitem{Lychagin} \emph{V.V. Lychagin,}\newblock Geometrical theory of
singularities of solutions to nonlinear differential equations, Problemi
Geometrii ( in Russian) VINITI, \textbf{20, }207-247 (1988).

\bibitem{Rakhimov} \emph{A.Kh. Rakhimov,}\newblock Singularities of Riemann
invariants, Func. Anal. Appl., \textbf{27} (1), 39-50 (1993).

\bibitem{Palamod} \emph{V.P. Palamodov,}\newblock Tangent fields on
deformations of complex spaces, \ Math. USSR-Sb., \textbf{71} (1), 163-182
(1992).

\bibitem{Witten1} \emph{E. Witten,}\newblock Topological quantum field
theory, Commun. Math. Phys., \textbf{117}, 353-386, (1988).

\bibitem{Witten2} \emph{E. Witten,}\newblock On the structure of topological
phase of two-dimensional gravity, Nucl. Phys., B \textbf{340}, 281-332,
(1990).

\bibitem{DVV} \emph{R.Dijkgraaf, H. Verlinde, E. Verlinde,}\newblock %
Topological strings in $d<1$, Nucl. Phys., B \textbf{352}, 59-86, (1991).

\bibitem{Saito} \emph{K. Saito,}\newblock Quasihomogeneisolierte
singulatitaten von hyperflachen, Invent. Math., \textbf{14,} 123-142, (1971).

\bibitem{Dubr1} \emph{B. Dubrovin,}\newblock\ Geometry of 2D topological
field theory, Lecture Notes in Math., \textbf{1620}, pp. 120-348, \
Springer-Verlag, Berlin, 1996.

\bibitem{Dubr2} \emph{B. Dubrovin,}\newblock Differential geometry of the
space of orbits of a Coxeter group, \ Survey in Diff. Geom., IV, (C.L.Terng,
K.Uhlenbeck, eds), Int. Press, Boston \ MA, 1998, pp. 181-211.

\bibitem{Krichever} \emph{I. Krichever,}\newblock The dispersionless Lax
equations and topological minimal models, Commun. Math. Phys., \textbf{143},
415-429 (1992).

\bibitem{Hert} \emph{C. Hertling,}\newblock Frobenius manifolds and moduli
spaces for singularities, Cambridge Univ. Press, Cambridge, 2002.

\bibitem{Givental2} \emph{A. Givental,}\newblock $A_{n-1}$ singularities and
nKDV hierarchies, Moscow Math. J., \textbf{3,} 475-505 (2003).

\bibitem{Giv-milanov} \emph{A. Givental, T. Milanov,}\newblock Simple
singularities and integrable hierarchies, Prog. Math., \textbf{232, }%
173-201, Birkhauser, Boston MA, 2005.

\bibitem{Fren-Giv-Milan} \emph{E. Frenkel, A. Givental, T. Milanov,}%
\newblock
Soliton equations, vertex operators and simple singularities, Func.Anal.
Other Math., \textbf{3, }47-63 (2010).

\bibitem{Kodama-Kon} \emph{Y. Kodama, B.G. Konopelchenko, }\newblock
Singular sector of the Burgers-Hopf hierarchy and deformations of
hyperelliptic curves, J. Phys. A: Math. Gen., \textbf{35,} L489-L500 (2002).

\bibitem{Dubr3} \emph{B. Dubrovin,}\newblock On Hamiltonian pertubations of
hyperbolic systems of conservation laws. II. Universality of critical
behaviour, Comm. Math. Phys., \textbf{267 }, 117-139 (2006).

\bibitem{Dubr4} \emph{B. Dubrovin,}\newblock On universality of critical
behaviour in Hamiltonian PDEs, Amer. Math. Soc. Transl. Ser. 2, \textbf{224}%
, 59-109, AMS, Providence RI, 2008.

\bibitem{Dubr-Grava-Klein} \emph{B. Dubrovin, T. Grava, C. Klein,}\newblock
On universality of critical behaviour in the focusing nonliner Schrodinger
equation, elliptic umbilic catastrophe and the tritronquee solution to the
Painleve-I equation, J. Nonlinear Sci., \textbf{19, }57-94 (2009).

\bibitem{Kon-Alonso-Med1} \emph{\ B.G. Konopelchenko, L. Martinez Alonso, E.
Medina,}\newblock Hodograph solutions of the dispersionless coupled KdV
hierarchies, critical points and the Euler-Poisson-Darboux equation, J.
Phys. A: Math. Gen., \textbf{43,} 434020 (2010).

\bibitem{Kon-Alonso-Med2} \emph{B.G. Konopelchenko, L. Martinez Alonso, E.
Medina,}\newblock Singular sectors of the one-layer Benney and
dispersionless Toda systems and their interrelations, Theor. Math. Phys., 
\textbf{168 }(1), 963-973 (2011).

\bibitem{Kon-Alonso-Med3} \emph{B.G. Konopelchenko, L. Martinez Alonso, E.
Medina,}\newblock On the singular sectors of the Hermitian random matrix
models in the large N limit, Phys. Lett.A,\textbf{\ 375, }867-872, (2011).

\bibitem{Whitham} \emph{G.B. Whitham,}\newblock Linear and nonlinear waves,
Wiley, New York, 1974.

\bibitem{Zakharov} \emph{V.E. Zakharov,}\newblock Benney equations and
quasiclassical approximation in the inverse problem method, Func.Anal.Appl., 
\textbf{14}, 89-98 (1980).

\bibitem{Rozd} \emph{B.L. Rozdestvenskii, N.N. Janenko,}\newblock Systems of
quasilinear equations and their applications to gas dynamics, Transl. Math.
Monographs, vol.55, AMS, Providence, RI, 1983.

\bibitem{Pavl} \emph{M.V. Pavlov,}\newblock Integrable hydrodynamic chains,
J. Math. Phys., \textbf{44}, 4134-4156 (2003).

\bibitem{Darboux} \emph{G. Darboux,}\newblock Lecons sur la theorie general
des surfaces, II, Gauthier Villars, 1915.

\bibitem{Tsarev} \emph{S.P. Tsarev,}\newblock The geometry of Hamiltonian
systems of hydrodynamic type. The generalized hodograph method, Math. USSR,
Izv., \textbf{37}, 397-419, (1991).

\bibitem{Arnold4} \emph{V.I. Arnold,}\newblock Local normal forms of
functions, Invent. Math., \textbf{35}, 87-109, (1976).

\bibitem{vander} \emph{B.L.van der Waerden,}\newblock Algebra,
Springer-Verlag, Berlin, 1967.
\end{thebibliography}
\end{document}